\documentclass[aps,showkeys]{revtex4-2}

\usepackage{amsfonts}
\usepackage{amsmath}
\usepackage{amssymb}
\usepackage{color}
\usepackage{graphicx}
\usepackage{multirow}
\usepackage{booktabs}
\usepackage[colorlinks=true,allcolors=blue]{hyperref}

\begin{document}

\title{Supersymmetric Partners of the One-Dimensional Infinite Square Well Hamiltonian}

\author{M. Gadella\footnote{ORCID: \href{http://orcid.org/00000-0001-8860-990X}{0000-0001-8860-990X}}}
\email{gadella@fta.uva.es}
\affiliation{Departamento de F\'{\i}sica Te\'{o}rica, At\'{o}mica y \'{O}ptica and IMUVA,
Universidad de Valladolid, 47011 Valladolid, Spain}

\author{J. Hern\'andez-Mu\~noz\footnote{ORCID: \href{http://orcid.org/0000-0002-0072-6077}{0000-0002-0072-6077}}}
\email{jose.hernandezm@uam.es}
\affiliation{Departamento  de F\'{\i}sica Te\'orica de la Materia Condensada, IFIMAC Condensed Matter Physics Center, Universidad Aut\'onoma de Madrid, 28049 Madrid, Spain}

\author{L. M. Nieto\footnote{ORCID: \href{http://orcid.org/0000-0002-2849-2647}{0000-0002-2849-2647}}}
\email{luismiguel.nieto.calzada@uva.es}
\affiliation{Departamento de F\'{\i}sica Te\'{o}rica, At\'{o}mica y \'{O}ptica and IMUVA,
Universidad de Valladolid, 47011 Valladolid, Spain}

\author{C. San Mill\'an\footnote{ORCID: \href{http://orcid.org/0000-0001-7506-5552}{0000-0001-7506-5552}}}
\email{carlos.san-millan@alumnos.uva.es}
\affiliation{Departamento de F\'{\i}sica Te\'{o}rica, At\'{o}mica y \'{O}ptica and IMUVA,
Universidad de Valladolid, 47011 Valladolid, Spain}

\date{\today}

\begin{abstract}
We find supersymmetric partners of a family of self-adjoint operators which are self-adjoint extensions of the differential operator $-d^2/dx^2$ on $L^2[-a,a]$, $a>0$, that is,  the one dimensional infinite square well. First of all, we classify these self-adjoint extensions in terms of several choices of the parameters determining each of the extensions. There are essentially two big groups of extensions. In one, the ground state has strictly positive energy. On the other, either the ground state has zero or negative energy. In the present paper, we show that each of the extensions belonging to the first group (energy of ground state strictly positive) has an infinite sequence of supersymmetric partners, such that the $\ell$-th order partner differs in one energy level from both the $(\ell-1)$-th and the $(\ell+1)$-th order partners. In general, the eigenvalues for each of the self-adjoint extensions of $-d^2/dx^2$ come from a transcendental equation and are all infinite. For the case under our study, we determine the eigenvalues, which are also infinite, {all the extensions have a purely discrete spectrum,} and their respective eigenfunctions for all of its $\ell$-th supersymmetric partners of each extension.
\end{abstract}


\maketitle

\section{Introduction}
\label{sec1}

The study of one dimensional models in quantum mechanics is useful in order to gain a better understanding of the properties of quantum systems. In~particular, the~construction of supersymmetric (SUSY) partners of given potentials allow for an analysis of one dimensional Hamiltonians that often keep similarities with the original ones. Many studies have been done in this field and a brief account of references~\cite{CHIS,INF,LRB,RRR,BAG,CKS,DNNR,MNR,FNN,INNN,CNP,GN,CJNP,FGN} only covers a small part of all previous~work.

In the present paper, we intend to investigate the properties of the SUSY partners of the self-adjoint determinations of the operator $-d^2/dx^2$ on $L^2[-a,a]$, $a>0$ and finite, with~appropriate boundary conditions at the points $-a$ and $a$. Note that this problem is closely related to the problem of the definition of the ``free'' Hamiltonian on the one dimensional infinite square well~potential. 

{From our point of view, SUSY quantum mechanics is a method that pursues the identification of the class of Hamiltonians for which their spectral problem can be algebraically solved. Traditionally, people have investigated SUSY partners of well studied exactly solvable Hamiltonians that give rise to other Hamiltonians for which the spectrum coincides with the spectrum of the original Hamiltonian except for one eigenvalue. In~addition, there are several examples in which one original Hamiltonian produces an infinite chain of Hamiltonians, the~first element of the chain being its SUSY partner and each of the others is a partner of the previous and the next one. Here, we explore the possibility of obtaining the whole chain of partners corresponding to self-adjoint extensions of a symmetric one dimensional Hamiltonian with equal deficiency indices. Since in our case, the~variety of self-adjoint extensions is quite wide, depending on four real parameters, we have expected to find interesting new results in the field as it happened to be.}

The analysis of these self-adjoint extension has been done in~\cite{BFV}. The~task of computing the SUSY partners of all the self-adjoint determinations (also called extensions) of  $-d^2/dx^2$ on $L^2[-a,a]$,  their spectra and their wave functions is not trivial, although~can be carried out~systematically. 

Although the idea of self-adjoint extensions of symmetric (or Hermitian) operators on (infinite dimensional) Hilbert spaces is not yet very popular among physicists, {it is, however,} possible to find recent papers on the topic~\cite{donaire,juanito1,juanito2,AFR,ZTZ,ZOL,ZOL1,GOL,EU}. Standard quantum mechanics textbooks refer to the one dimensional infinite square well potential or the harmonic oscillator as if they were described by a unique self-adjoint Hamiltonian, which produces a neatly calculable  spectrum. The~mathematical reality is much more complex and may give many more possibilities for the study of quantum mechanics systems. Let us briefly address to this problem, for~which a more {thorough} presentation can be found in mathematical textbooks~\cite{RSII} as well as papers addressed to the Physics community~\cite{BFV}.

Concerning terminology, an~operator, $A$, on~a infinitely dimensional separable Hilbert 
space $\mathcal H$  (the Hilbert space must be infinite dimensional, since otherwise all operators are continuous and defined on the whole space. In~such a case, this argumentation does not make sense. A~separable Hilbert space is one with a countable orthonormal basis, which is always the case in ordinary quantum mechanics) is symmetric, or~equivalently Hermitian if~for any pair of vectors $\varphi,\psi\in\mathcal D(A)$, where $\mathcal D(A)$ is the domain of $A$, which must be 
densely defined, one has that $\langle A\varphi|\psi\rangle=\langle\varphi|A\psi\rangle$, where $\langle-|-\rangle$ denotes the scalar product on $\mathcal H$. This means that the adjoint, $A^\dagger$, of~$A$ extends $A$, $A\prec A^\dagger$ (i.e., $\mathcal D(A) \subset \mathcal D(A^\dagger)$ and $A\psi=A^\dagger\psi$, for~all $\psi\in\mathcal D(A)$). The~deficiency indices are $n_\pm:=\dim \,{\rm Ran}(A^\dagger\pm iI) $, where ${\rm Ran}(B)$ is the range (image space) of the operator $B$ and $I$ is the identity operator.   A~symmetric (or Hermitian) operator has {self-adjoint} determinations (or extensions) if and only if $n_+ =n_-$ \cite{RSII}. If~$n_+=n_-=0$, this extension is unique. On~the other hand, if~$n_+ =n_- \ne 0$, the~number of extensions is infinite and, in~the case of Hilbert spaces of functions, they usually can be determined by some matching or boundary conditions that the functions in the domain of the extensions should {fulfill} at some points~\cite{BFV,RSII,fassari,KU}.

Self-adjoint determinations of the operator $-d^2/dx^2$ defined on functions supporting whatever interval, $\mathcal K$, in~the real line $\mathbb R$ are used to define the so call {\it {contact potentials}}~\mbox{\cite{greendelta,gaussian,fassari,romaniega,albeverio}}. These are perturbations of the ``free operator'' $H_0=-d^2/dx^2$, which are supported on a single point $x_0\in\mathcal K$. Typical examples of contact potentials are the Dirac delta $\delta(x-x_0)$ or its derivative $\delta'(x-x_0)$, which define Hamiltonians of the type $H_0+\delta(x-x_0)$ or $H_0+\delta'(x-x_0)$ as well defined self-adjoint operators on the Hilbert space $L^2(\mathcal K)$ \cite{KU}. These types of perturbations may serve as a good and tractable approximation for a very localized spatial perturbation and are defined via matching conditions that must satisfy the functions on the domain of the operator at  $x_0$.  Concerning the operator $-d^2/dx^2$ on $L^2[-a,a]$, some relations have been found among the boundary conditions at the borders $-a$ and $a$ and matching conditions defining a $\delta$ or $\delta'$ perturbation at the origin~\cite{GGGM,GGN}. 

A comment is of relevance here. Let us consider the subspace, $\mathcal D_0$, of~all twice differentiable square integrable functions, $\varphi(x)$, in~the interval $[-a,a]$, with~second derivative in $L^2[-a,a]$, verifying  the boundary conditions $\varphi(-a)=\varphi(a)=\varphi'(-a)=\varphi'(a)=0$, and~a differential operator of the form
\begin{equation}\label{1}
    D=-\dfrac{d^2}{dx^2}+p_1(x)\dfrac{d}{dx}+p_2(x),
\end{equation}
where $p_1(x)$ and $p_2(x)$ are continuous real functions (with $p_1(x)$ differentiable) on $[-a,a]$. 
Then $D$ is Hermitian on $\mathcal{D}_0$ with deficiency indices  {(2, 2).} 
 It has been proven in~\cite{NAI} (vol.~2, p. 90) that all self-adjoint extensions of $D$ have a purely discrete spectrum. This is precisely the case of all the self-adjoint determinations of $-d^2/dx^2$ under our study~\cite{BFV}. These self-adjoint extensions are characterized by a set of {four} real parameters, so that one particular choice of these parameters  gives a unique self-adjoint determination of $-d^2/dx^2$ on $L^2[-a,a]$ and vice-versa. Although~this is much less known, a~similar situation emerges in the study of the one dimensional harmonic oscillator~\cite{GGN1}. 

The present article intends in the first place, to complete as far as possible, the~classification of the self-adjoint extensions of $-d^2/dx^2$ on $L^2[-a,a]$ given by~\cite{BFV}. 
Once this task has been done, we intend to obtain the whole chain of SUSY partners of each of the self-adjoint extensions using standard methods already developed in the theory~\cite{CHIS}. This kind of supersymmetry intends to construct a series of potentials (in our case one-dimensional), with~an energy spectrum closely related and that can be obtained from the spectrum of the original potential. Thus, being given one of our original self-adjoint extensions and being known the solution of the spectral problem, we should be able to obtain an infinite sequence of Hamiltonians such that their spectra coincides with the spectra of the previous one except for one eigenvalue, and~hence from the original one except for a finite number of energy levels. We must add that all self-adjoint extensions of $-d^2/dx^2$ on $L^2[-a,a]$ have a purely discreet spectrum with an infinite number of energy~levels.

The ground state for each of these extensions either has a strictly positive, zero or negative energy. Obviously, in~the latter case, this fact comes from extensions which are not definitely positive. This is somehow paradoxical, due to the form of the original operator, which is $-d^2/dx^2$. This paradox is solved in~\cite{BFV}. For~those extensions with a ground state with strictly positive energy, we have constructed the whole sequence of its SUSY partners and have given the eigenvalues and eigenfunctions for these partners. As~mentioned earlier, the~set of eigenvalues for each partner comes from the set of eigenvalues of the extension from which we construct the sequence of~partners.

The general formalism can also be applied to obtain a sequence of Hamiltonians when the ground state of the original self-adjoint extension of $-d^2/dx^2$ on $L^2[-a,a]$ has zero or negative energy. In~this case, partner Hamiltonians may be very different from the original one in the sense that they may have a finite number of eigenvalues or simply no eigenvalues. This is due to the presence of nodes in the wave function of the ground state. Nevertheless, these partners may be obtained and classified, although~this discussion is left for a future~publication.

This paper is organized as follows---in Section~\ref{sec2} we reformulate the classification given by~\cite{BFV} of the self-adjoint extensions of $-d^2/dx^2$ on $L^2[-a,a]$.   In Section~\ref{sec3}, we classify  these extensions in terms of some other sets of parameters, not considered in~\cite{BFV}. In {Section}~\ref{sec4}, we construct the first SUSY partners for those extensions with positive ground level energy and give the precise form of its eigenfunctions. In {Section}~\ref{sec5}, we give the complete sequence of SUSY partners for each of these extensions. We close this article with a Conclusions Section  and an Appendix in which we show what the correct form for the wave functions for the energy levels should~be.

\section{Self-Adjoint Extensions: Determination of Their~Eigenvalues}
\label{sec2}

Let us go back to the differential operator $H_0:=-d^2/dx^2$ defined on $L^2[-a,a]$, $a>0$ and with domain $\mathcal D_0$ as above, just before \eqref{1}. On~$\mathcal D_0$, $H_0$ is symmetric (Hermitian) with deficiency indices (2, 2) \cite{BFV}. According to the von-Neumann theorem~\cite{RSII}, $H_0$ admits an infinite number of self-adjoint extensions labeled by four real~parameters.

The adjoint operator $H_0^\dagger$ acts as $-d^2/dx^2$ on the functions of its domain (see~\cite{RSI,BN} for a definition of the domain of the adjoint of a given densely defined operator and its properties). If~$\phi$ is a function of such domain, we get integrating by parts:
\begin{equation}
\left\langle-\dfrac{d^2}{dx^2}\phi,\phi \right\rangle=B(\phi,\phi)+ \left\langle \phi,-\dfrac{d^2}{dx^2}\phi \right\rangle \,,
\end{equation}
where $\langle -,-\rangle$ denotes the scalar product on $L^2[-a,a]$ and
\begin{eqnarray}\label{3}
 B(\phi,\phi)=\phi'(a)\phi^*(a)-\phi(a)\phi'^*(a)-\phi'(-a)\phi^*(-a)+\phi(-a)\phi'^*(-a),
    \end{eqnarray}
the prime being the derivative with respect to the variable $x$ and the asterisk meaning complex conjugate.
The self-adjoint extensions of $H_0$ are equal to $-d^2/dx^2$ as an operator acting on the subdomains of the domain of $H_0^\dagger$ of functions with $B(\phi,\phi)=0$. This happens if and only if there exists a $2\times2$ unitary matrix $U$ such that (see~\cite{BFV} and references quoted~therein):
\begin{equation}\label{4}
    \left(\begin{matrix}2a\phi'(-a)-i\phi(-a)\\[2ex]
    2a\phi'(a)+i\phi(a)\end{matrix}\right)=U\left(\begin{matrix}2a\phi'(-a)+i\phi(-a)\\[2ex] 2a\phi'(a)-i\phi(a)\end{matrix}\right).
\end{equation}

The set of self-adjoint extensions of $H_0$ is in one to one correspondence with the set of $2\times 2$ unitary operators $U$. Thus, each of these extensions will be labeled by its corresponding operator as $H_\alpha$. Since there is a set of four real independent parameters that characterize the set of operators $U$, then, the~set of self-adjoint extensions of $H_\alpha$ is also characterized by the same parameters~\cite{BFV}. Each of the operators $U$ has the following form~\cite{BFV}:
\begin{equation}\label{5}
    U=e^{i\psi}\left(\begin{matrix}
    m_0-im_3& -m_2-im_1\\[2ex]
    m_2-im_1& m_0+im_3
    \end{matrix}\right) \,.
\end{equation}

Here, $\psi$ and $m_i$, $i = 0,1,2,3$ are real parameters so that $\psi\in[0,\pi]$ and $m^2_0 + m^2_1 + m^2_2 + m^2_3 = 1$, which means that only four parameters are independent~\cite{BFV}. The~latter relation is a consequence of unitarity: the modulus of the determinant of $U$ must be a number of modulus~one.

There are some of these extensions with a clear physical interest, which does not mean that the others are irrelevant from the physics point of view. In~\cite{BFV}, the~authors distinguish three categories of~extensions: 
\begin{itemize}
\item[(i)]
Those which preserve time reversal; 
\item[(ii)] 
Those which preserve parity; 
\item[(iii)] 
Those preserving positivity. 
\end{itemize}

Apart from these three categories, there are some other extensions.  The~reason why the authors of~\cite{BFV} single out those  {extensions} that preserve positivity is due to the existence of extensions with negative energies. In~fact, as~proven in~\cite{NAI} (Theorem 16, vol 2, page 44), $H_\alpha$ may have one (which may be doubly degenerate) or two (with no degeneration) negative energy states.  All other extensions have non-negative eigenvalues and are called {\it positivity preserving}. Only three of {these} positivity preserving extensions with special simplicity are discussed in~\cite{BFV}. We want to determine the energy levels in this~situation. 
 
In order to obtain the energy levels for a specific self-adjoint extension, $H_\alpha$, of \mbox{$H_0=-d^2/dx^2$} on $L^2[-a,a]$, we have to solve the Schr\"odinger equation and impose on its solutions the boundary conditions that characterize the extension. These boundary conditions are given by the \eqref{4} and \eqref{6}. However as stated in~\cite{BFV}, the~determination of which operators $U$ satisfy the positivity condition as stated before involves tedious considerations.
To circumvent this difficulty, let us consider the general solution of the time independent Scr\"odinger equation $-d^2\phi(x)/dx^2=E\phi(x)$, with~$E=s^2/(2a)^2\ge 0$, where $2a$ is the infinite square well width {(Although the energy is given, in~our notation, by~$\hbar^{2}E/2m$, we are calling ``energy'' the quantity represented by $E$.)}. This general solution~is
\begin{equation}\label{6}
    \phi(x)=A \cos\left(\frac{sx}{2a}\right)+B\sin\left(\frac{sx}{2a}\right).
\end{equation}

Here, $A$ and $B$ have to be fixed with two conditions: (i) $\phi(x)$ should be normalized in $L^2[-a,a]$ and (ii) $\phi(x)$ should fulfill the boundary conditions \eqref{4} and \eqref{5} so that $E\ge 0$. Let us use \eqref{6} in relation \eqref{4} giving the general matching conditions, so as to obtain the following homogeneous linear system:
\begin{equation}\label{7}
\Bigl( \mathcal{L}(s)-U \mathcal{M}(s)\Bigr)\left(\begin{matrix}A\\ B\end{matrix}\right)= \mathcal{N}(s)\left(\begin{matrix}A\\ B\end{matrix}\right)=0\,,
\end{equation}
where
\begin{eqnarray}
\mathcal{L}(s)=\left(\!\!\!\begin{array}{cc}
s \sin \frac{s}{2} -i \cos\frac{s}{2}  & s \cos\frac{s}{2} +i \sin\frac{s}{2} \\ [1ex]
 -s \sin\frac{s}{2}+i \cos\frac{s}{2} & s \cos\frac{s}{2}+i \sin\frac{s}{2}
 \end{array}\!\!\right), \qquad 
\mathcal{M}(s)=\left(\!\!\!\begin{array}{cc}
s \sin\frac{s}{2} +i \cos\frac{s}{2}  & s \cos\frac{s}{2}-i \sin\frac{s}{2} \\ [1ex]
-s \sin(\frac{s}{2}-i \cos\frac{s}{2} & s \cos\frac{s}{2}\-i \sin\frac{s}{2} 
 \end{array}\!\!\right).
\end{eqnarray}
The eigenvalues $\lambda_\pm(s)$ of the matrix  $\mathcal{N}(s)$ are given by
\begin{equation}\label{10}
    \lambda_\pm(s)=\frac{\operatorname{Tr}(\mathcal{N}(s))}{2}\pm\sqrt{\left(\frac{\operatorname{Tr}(\mathcal{N}(s))}{2}\right)^2-\det(\mathcal{N}(s))}.
\end{equation}

The trace and the determinant of $\mathcal{N}(s)$ can be easily calculated and are, respectively:
\begin{equation}\label{11}
{\rm Tr}(\mathcal{N}(s))=e^{-\frac{1}{2} i (s-2 \psi )} \left(-m_3 (s+1)+i m_2 e^{i s} (s-1)\right)\,,
\end{equation}
and
\begin{equation}\label{12}
\det(\mathcal{N}(s))=-4ie^{i\psi}\left[(m_0+\cos\psi)\sin s+2s(m_1-\cos s\sin\psi)-s^2(m_0-\cos\psi)\sin s \right]\,.
\end{equation}

To begin with, let us remark that in order to have non-trivial solutions of \eqref{7} we must~have
\begin{equation}\label{detn0}
\det\left(\mathcal{N}(s)\right)=0.
\end{equation}

Then, the~set of eigenvalues of $\mathcal{N}(s)$ is given by $\operatorname{Tr}(\mathcal{N}(s))$ and  $0$, as~may be immediately seen from \eqref{10}. The~condition \eqref{detn0} gives a relation between the values of the energy, determined by the real parameter $s$, since $E=s^2/(2a)^2$, and~the parameters $\psi$, $m_0$ and $m_1$, as~in \eqref{5}. In~consequence, the~energy levels depend on the values of these three parameters only. From~  \eqref{12} and \eqref{detn0} , we obtain the following two transcendental equations (one with plus sign and the other with minus sign):
\begin{equation}\label{13}
   s\,\sin s= \frac{m_1-\cos s \sin\psi}{m_0-\cos\psi}\pm\sqrt{\left(\frac{m_1-\cos s \sin\psi}{m_0-\cos\psi}\right)^2+\frac{ m_0+\cos\psi}{m_0-\cos\psi}\, \sin^2s}\,.
\end{equation}

This form of the transcendental equations is quite interesting, since it will serve as an efficient estimation of the energy levels when these values cannot be exactly calculated. Otherwise, they permit to obtain exact solutions whenever they exists. 
 Let us now summarize three of the results provided by~\cite{BFV}, which we will need later~on:
\begin{itemize}
\item
The eigenvector $(A,B)$ of $\mathcal{N}(s)$ with $0$ eigenvalue is given by
{\small \begin{eqnarray}
 \label{A14}
\hskip-2cm A\!&\!=\!&\!
\left[ i+e^{i \psi } (i m_0+m_1-i m_2+m_3)\right] \sin\frac{s}{2}  +s\left[ 1+ e^{i \psi } (m_0+i m_1+m_2+i m_3)\right] \cos\frac{s}{2} ,
\\ [2ex] 
\label{B15}
\hskip-2cm B\!&\!=\!&\! s \left[-1+e^{i \psi } (m_0+i m_1+m_2-i m_3)\right] \sin\frac{s}{2} +\left[ i+ e^{i \psi } (i m_0-m_1+i m_2+m_3)\right]\cos\frac{s}{2}.
\end{eqnarray}}
These expressions generalize similar ones published in citation [14] of our Reference~\cite{BFV}.
We see that the eigenvector depends on all the parameters $(m_0,m_1,m_2,m_3,\psi)$.

\item
 The extensions preserving time reversal invariance, are given by
\begin{equation}\label{16}
         m_2=0\,.
      \end{equation}     
\item
The parity preserving extensions of $H_0$ are those for which the eigenfunctions $\phi(x)$~verify :
\begin{equation}\label{17}
|\phi(x)|^2 =|\phi(-x)|^2 \Longrightarrow    |\phi(a)|^2=|\phi(-a)|^2\,.
\end{equation}
\end{itemize}

\subsection{Parity Preserving Extensions of $H_0$}
We are interested now in getting more information on the parity preserving extensions of $H_0$. Then, if~we use \eqref{6} in \eqref{17}  we obtain that $\operatorname{Re}(A\,B^*)\, \sin s=0$.
Hence, either
\begin{equation}\label{re19}
\sin s=0 \quad \text{or}\quad  \operatorname{Re}(A\,B^*)=0 .
\end{equation}

{Compare to Equation~(68) in citation {\cite{FGN}} of our Reference~\cite{BFV}.}
Taking into account the values of $(A,B)$ given in \eqref{A14} and \eqref{B15}  and also the fact that  $\det(\mathcal{N}(s))=0$, the~second equation of \eqref{re19} implies that either $m_3=0$ 
or
\begin{equation}\label{15}
(m_3+\sin\psi)\sin s+2s (m_2+\cos s \cos\psi)+s^2(\sin\psi -m_3)\sin s  =0.
\end{equation}

Solving this equation {\it {as if it were a quadratic equation on $s$}} gives a pair of transcendental equations which closely resemble Equation \eqref{13}. Thus, the~complete set of solutions of \eqref{re19}~are
\begin{subequations}
\begin{gather}
    m_3=0\,, \label{21a}\\      
    \sin s=0\,, \label{21b}\\ 
   s\,\sin s=\frac{m_2+\cos s \cos\psi}{m_3-\sin\psi}\pm \sqrt{\left(\frac{m_2+\cos s \cos\psi}{m_3-\sin\psi}\right)^2 +\frac{m_3+\sin\psi}{m_3-\sin\psi}\, \sin^2s}\,.  \label{21c}\end{gather}  
\end{subequations}  

Hence, when the parity is preserved, Equation \eqref{13} holds. This happens for three different situations given by  Equations \eqref{21a}--\eqref{21c}. These formulas, plus \eqref{13}, which derives from such a general principle as  $\det(\mathcal{N}(s))=0$, should give the energy levels for the infinite square well with parity preserving self-adjoint extensions, $H_\alpha$, of~$H_0$. 

Equation \eqref{21a} does not provide any extra information, \eqref{13} being the only relation which gives information on the energy spectrum. This parity preserving condition $m_3=0$ has been already used in~\cite{BFV}, although~in this paper relations \eqref{21b} and \eqref{21c} are not~mentioned. 

Equation \eqref{21b} obviously gives an energy spectrum of the parity preserving extensions that coincides to the spectrum given by texts in Quantum Mechanics for the extension with domain given by functions with $\phi(-a) = \phi(a) =0$. Henceforth, we shall call this extension the {\it textbook extension}. 

 Finally, \eqref{21c} gives the energy levels for other parity preserving extensions in terms of the three parameters $(\psi,m_2,m_3)$.

  In consequence, we have eight different situations for those extension having a non-negative spectrum, including those with time reversal and parity invariance, as~shown in Table~\ref{table:1}. In~the next section we will analyze some of these~situations.

\begin{table}[htb]
\centering
\caption{List of how to obtain the possible spectra as a function of the conserved~properties.}
\begin{tabular}{c c c} 
 \toprule
 \multicolumn{3}{c}{\textbf{Generic Spectrum:} \eqref{13}}\\[0.5ex]
 \midrule 
 \multicolumn{3}{c}{Time reversal invariance:  \eqref{13} and \eqref{16}, or~$m_2=0$}\\[0.5ex]
\midrule 
\multirow{7.5}{2.9 cm}{Parity preserving:} & \multirow{2.5}{2.2cm}{\eqref{13} and \eqref{21a}} & $m_2=0$\\ 
          \cmidrule{3-3}
          & &$m_2\neq0$\\
         \cmidrule{2-3}
       &   \multirow{2.5}{2.2cm}{\eqref{13} and  \eqref{21b}} &$m_2=0$\\ 
          \cmidrule{3-3}
         & &$m_2\neq0$\\
           \cmidrule{2-3}
        &  \multirow{2.5}{2.2cm}{\eqref{13} and \eqref{21c}} & $m_2=0$\\
           \cmidrule{3-3}
                & &$m_2\neq0$\\
 \bottomrule
\end{tabular}
\label{table:1}
\end{table}
\unskip

\section{Spectrum of the Free Particle on a Finite~Interval}
\label{sec3}

One of the goals of our study is to solve the eigenvalue problem for all the self-adjoint extensions, $H_\alpha$, of~the operator $H_0=-d^2/dx^2$ on $L^2[-a,a]$, which from the point of view of the physicist is the infinite square well with width $2a$. As~we have already seen, there are only a few of these extensions for which we may obtain an exact solution, including the {\it {textbook extension}}. For~most of these extensions the energy levels are solutions of a transcendental equation and, therefore, no explicit solutions of the eigenvalue problem for these extensions can be~given.

\subsection{The Angular Representation of the Self-Adjoint Extensions of $H_0$}

Due to the relation between the parameters $m_i$, given by
\begin{equation}\label{eqemes}
m_0^2+m_1^2+m_2^2+m_3^2=1, 
\end{equation}
a new parametric representation of the self-adjoint extensions, $H_\alpha$, of~$H_0=-d^2/dx^2$ on $L^2[-a,a]$ in terms of angular variables only is possible. Apart from the variable $\psi$, which is already angular, so that we keep it untouched, we have three other angular variables, $\theta_i$, $i=0,1,2$, defined by means of the following relations:
\begin{equation}\label{18}
m_0=\cos\theta_1\cos\theta_0, \quad m_1=\cos\theta_1\sin\theta_0, \quad m_2=\sin\theta_1\cos\theta_2, \quad m_3=\sin\theta_1\sin\theta_2.
\end{equation}

\begin{figure}[htb]
        \includegraphics[width=0.9\textwidth]{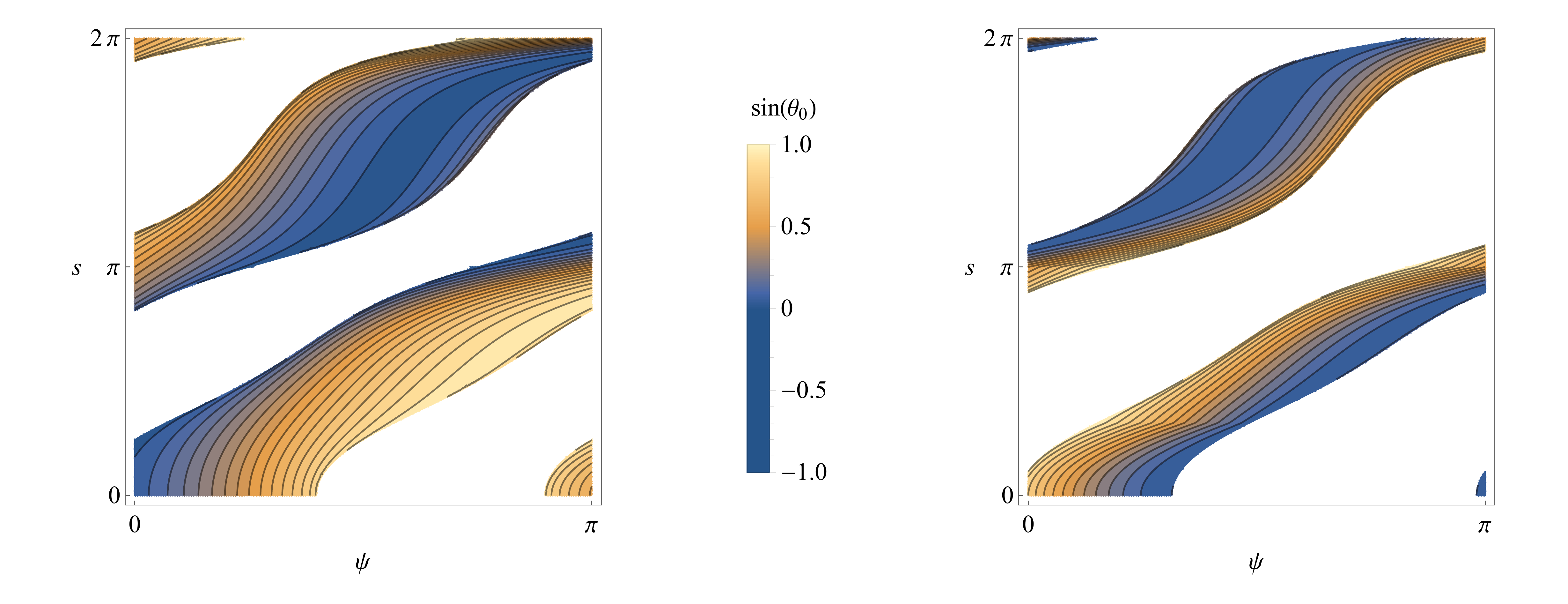}
\caption{Two plots of the implicit Equation \eqref{13} with the parametrization \eqref{18} allow us to see the variation of the parameter $s$ (remember that $E=s^2/(2a)^2$) as a function of $\psi$ and $\sin\theta_0$: on the left for $\theta_1=\pi/4$, on~the right for $\theta_1=4\pi/3$.}
\label{fig:fig1}
\end{figure}

Taking into account that Equation \eqref{13} gives the values of $s$, and~hence the energy levels, in~terms of the triplet of parameters $(\psi,m_0,m_1)$, then, according to \eqref{18},  $s$ will depend on the angular variables $(\psi,\theta_0,\theta_1)$ only.  
In general, we cannot solve \eqref{13} to find $s(\psi,\theta_0,\theta_1)$ explicitly. Since \eqref{13} depends on four  parameters $(s,\psi,\theta_0,\theta_1)$, we cannot represent  this equation in general but, as~in Figure~\ref{fig:fig1}, we can plot the square root of the energy (essentially $s$) for given values of $\theta_1$ and $\sin\theta_0$ as a function of $\psi$.

The general case can neither be explicitly solved nor represented graphically. Yet, there are two other situations sharing this negative characteristics. One is $m_2=0$ (time invariance only) and $m_3=0$ (parity conservation only). All other cases either can be explicitly or graphically solved or~both.

\subsection{Some Simple~Cases}
In the sequel we are going to deal with the cases of Table~\ref{table:1} that can be treated in some way, either graphically or~analytically.

\subsubsection{Parity and Time Reversal Invariance: $m_2=m_3=0$}

\begin{figure}[htb]
   \includegraphics[width=0.5\textwidth]{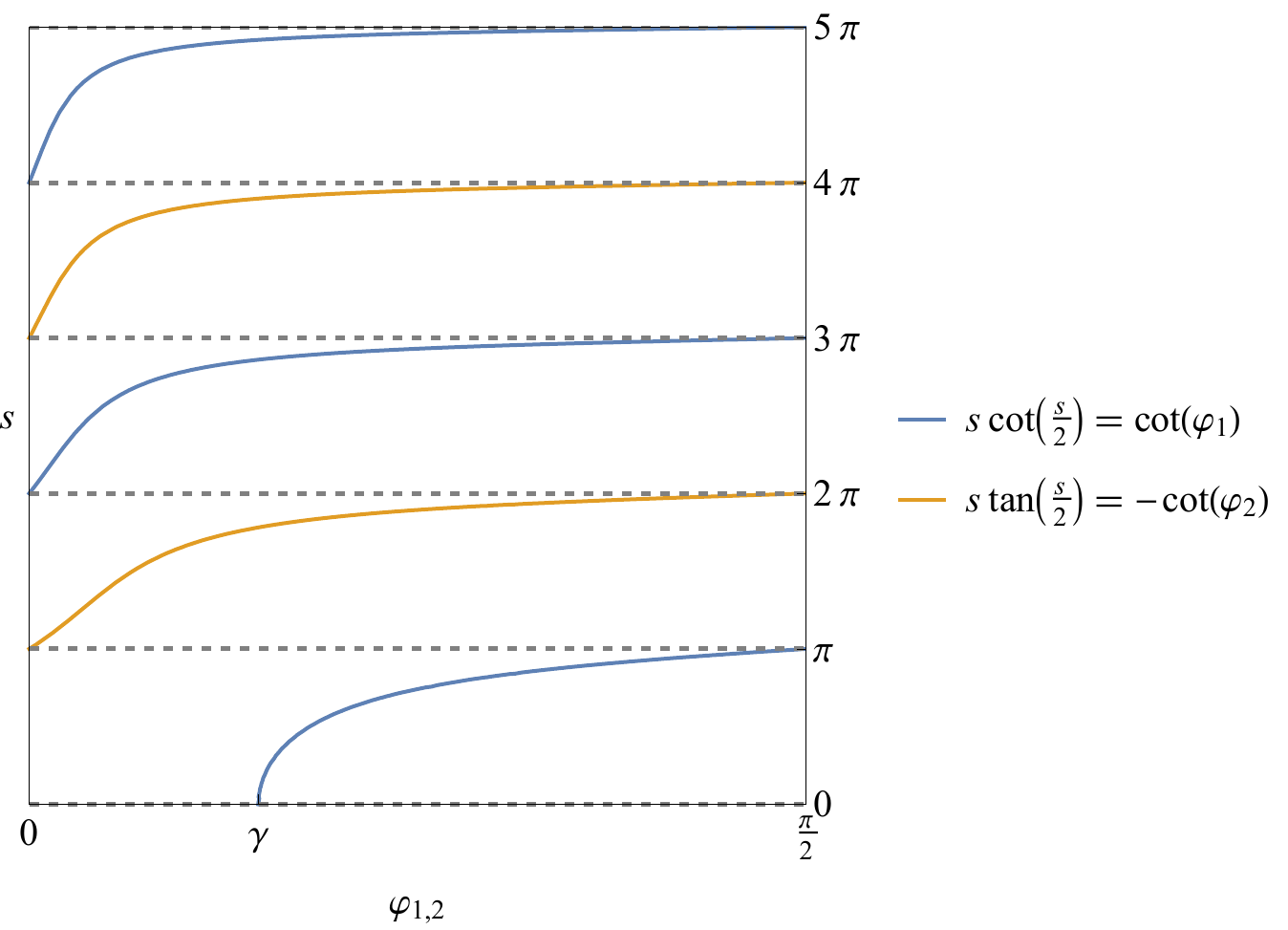}
\caption{Energy levels ($E=s^2/(2a)^2$) for odd parity  (blue) and  even parity  (yellow) solution for $m_2=m_3=0$, coming from  \eqref{19}.}
\label{fig:fig2}
\end{figure}

For $m_2=m_3=0$, we have in \eqref{eqemes} that $m_0^2 + m_1^2=1$, that is, we can take $\theta_1=0$ in \eqref{18} and therefore $m_0= \cos \theta_0$ and $m_1 = \sin\theta_0$. 
Then, let us go back to \eqref{11}, so as to see that $\text{Tr}(\mathcal{N}(s))=0$. Since one of the eigenvalues of $\mathcal{N}(s)$ must be zero, the~fact that $\text{Tr}(\mathcal{N}(s))=0$ makes the second eigenvalue also equal to zero. Thus, the~matrix $\mathcal{N}(s)$ admits a Jordan decomposition in terms of an upper triangular matrix. Now, the~transcendental Equation~\eqref{13} becomes much simpler, still depending on the sign, $\pm$, of~the square root. Note that this sign is positive if $s\in\{(0,\pi),(2\pi,3\pi)..\}$ and negative if $s\in\{(\pi,2\pi),(3\pi,4\pi),..\}$.  In~these two situations, the~spectral equation, the~vector $(A,B)$ and the eigenfunctions have the following explicit forms:
\begin{equation}\label{19}
    \begin{cases}
    \text{Positive square root} &  \begin{cases}
    \text{Spectrum equation:} &  s\tan\left(\frac{s}{2}\right)=-\cot\left(\frac{\psi+\theta_0}{2}\right)\\
    \text{Eigenvector:} &(A,B)= (1,0)\\
    \text{Eigenfunction:}& \phi(x)=\cos\left(\frac{s_0x}{2a}\right)
    \end{cases}\\ \\
        \text{Negative  square root} &  \begin{cases}
    \text{Spectrum equation:} &  s\cot\left(\frac{s}{2}\right)=\cot\left(\frac{\psi-\theta_0}{2}\right)\\
    \text{Eigenvector:} &  (A,B)=(0,1)\\
    \text{Eigenfunction:}& \phi(x)=\sin\left(\frac{s_0x}{2a}\right).
    \end{cases} \\
    \end{cases}
\end{equation}
In the above expressions for the spectral equations, we may write $\frac{\psi-\theta_0}{2}=\varphi_1$ and $\frac{\psi+\theta_0}{2}=\varphi_2$, where $\varphi_i$, $i=1,2$ are two {independent} angles.  Both spectral equations are represented in Figure~\ref{fig:fig2}.
The combination of both solutions tend to the textbook solution in the limit $\varphi_{1,2} \to 0$ for both angles. The~ground state for the textbook solution comes from the lowest state for the even parity preserving extensions. It is remarkable that if $\varphi_1\geq \arctan(1/2):=\gamma$, then the ground state  no longer comes from the even but from the odd parity preserving extensions, as~can be clearly seen in Figure~\ref{fig:fig2}.

\unskip

\subsubsection{Parity Preserving Extensions Fulfilling $\sin s=0$}

In this situation, Equation \eqref{21b} implies that $s=n\pi$ with $n=0,\pm1, \pm 2,\dots$, so that $E=(n^2\pi^2)/(2a)^2$, $n=1,2,\dots$.  The~energy levels of all these extensions are the same as in the textbook's extension. No negative energy states may exist. In~order to obtain the corresponding eigenfunctions, which may be different from those obtained for the textbook case, we parametrize these extensions using angular variables. However, we are now using a different angle parameterization from \eqref{18}. Indeed, taking into account \eqref{13}, we will find it very useful to use this one:
\begin{eqnarray}\label{20}
m_0=\cos\psi\, \cos\beta_0,  &   m_1=(-1)^n \sin\psi,  \\ [1ex]  
m_2=\cos\psi\, \sin\beta_0\, \cos\beta_1,  &   m_3=\cos\psi\, \sin\beta_0\, \sin\beta_1,
\label{20a}
\end{eqnarray}
where the exponent $n$ appearing in the expression for $m_1$ is the same number that labels $s$. 

The eigenfunctions must obey the Schr\"odinger equation, so that they should be of the form \eqref{6}. The~coefficients $A$ and $B$ depend on the energy levels and, therefore, should be functions of $s$. In~the simple case in which $s$ be a even or odd multiple of $2\pi$, these coefficients can be obtained from the following relations,  using \eqref{A14} and \eqref{B15} and the new~parameterization:
\begin{equation}\label{26}
    \begin{cases}
    s=2q\pi:& \begin{cases}
    A(s)=2 \pi   q \left(1+ e^{i \beta_1} \sin \beta_0-\cos \beta_0\right),\\
       B(s)=i   \left(1+e^{-i \beta_1} \sin \beta_0+\cos \beta_0\right),
    \end{cases} q=1,2,\dots\\[3ex]
     s=\left(2q+1\right)\pi: & \begin{cases}
    A(s)=i \left(1-e^{i \beta_1} \sin \beta_0+\cos \beta_0\right),\\
        B(s)=\pi   (2 q+1)  \left(-1+e^{-i \beta_1} \sin \beta_0 +\cos\beta_0\right),
    \end{cases}  q=0,1,\dots
    \end{cases}
\end{equation}

The eigenfunctions depend on the angles $(\psi,\beta_0,\beta_1)$ only. These angles are those defined in \eqref{20} and \eqref{20a}; note the difference with the angles defined in \eqref{18}. 
  If we take the limits $\beta_0\to 0$ and $\beta_1\to 0$, we recover the eigenfunctions for the textbook~extension.

\subsubsection{Parity and Time Reversal Invariance Extensions Fulfilling (\ref{21c})}

We have seen that \eqref{13} has a general validity, which is independent of the particular situation under study. On~the other hand, \eqref{21c} is valid for extensions that preserve parity invariance. Note that the left hand side in both equations is the same: $s\sin(s)$. This suggests that the identity between both right hand sides would help to solve the spectral equation in this case. This identity gives:
\begin{eqnarray}\label{277}
&& \frac{m_1-\cos s \sin\psi}{m_0-\cos \psi}\pm\sqrt{\left(\frac{m_1-\cos s \sin\psi}{m_0-\cos \psi}\right)^2+\frac{ m_0+\cos \psi }{m_0-\cos\psi}\sin^2s}
\nonumber\\
&&\qquad =
\frac{m_2+\cos s \cos\psi}{m_3-\sin\psi}\pm \sqrt{\left(\frac{m_2+\cos s\cos\psi}{m_3-\sin\psi}\right)^2 +\frac{m_3+\sin\psi}{m_3-\sin\psi}\sin^2s}\,,
\end{eqnarray}
which may be written in polynomial form as $\cos^4s+P_1\cos^3s+P_2\cos^2s +P_3 \cos s+P_4=0$, where the functions $P_i$ depend on the parameters $(m_0,m_1,m_2,m_3,\psi)$. We do not write the precise form of this polynomial relation in here, since it is extremely long and it does not show interesting features. Nevertheless, it is important to note that this is a fourth order polynomial on the variable $\cos s$ with coefficients depending on the parameters. Two of these solutions of \eqref{277} are
\begin{equation}\label{28}
    \cos s=\pm 1\,.
\end{equation}

These solutions may be written as $\sin s=0$, which coincide with \eqref{21b}, so that no new solutions for the spectral problem arise from \eqref{28}. The~other two solutions are quadratic as function of the parameters and are rather huge and intractable. To~simplify this problem as much as possible, let us define a third and last angle re-parameterization of the $m_i$:
\begin{equation}\label{29}
m_0=\sin\omega_1 \cos\omega_2\,, \quad m_1=\cos\omega_1\sin\omega_0\,, \quad m_2=\cos\omega_1\cos\omega_0\,, \quad m_3=\sin\omega_1\sin\omega_2\,.
\end{equation}

This parameterization is quite similar to \eqref{18}, where we have interchanged the expressions for $m_0$ and $m_2$. In~terms of the new angular variables, an~expression for the energy levels as functions of $s$ is given by
\begin{eqnarray}\label{300}
&& \cos s=-\cos \omega_1 \cos (\omega_0+\omega_2) \cos (\omega_2-\psi )    
\\[2ex] 
&&\qquad\qquad    - \sec\omega_1 \left[\sin
   (\omega_0+\omega_2) \sin (\omega_2-\psi )  
   \pm i     \sqrt{\sin^4\omega_1 \cos ^2(\omega_0+\omega_2) \sin ^2(\omega_2-\psi )}\, \right]. \nonumber
\end{eqnarray}

As we want $s$ to be real (in order to have positive eigenvalues of the energy), the~imaginary term in \eqref{300} must vanish. Note that all factors under the square root are positive, so that the eigenvalues of the energy can be found, with all those in Equation \eqref{300} for each of the factors under the square root vanishing. 
There are three possibilities, which yield the following equations:
\begin{subequations}
\begin{gather}
\sin\omega_1=0 \implies \cos s=\pm \cos(\omega_0+\psi)\implies  s=n\pi\pm(\omega_0+\psi)\,, \ \ \label{311a} 
\\
\cos(\omega_0+ \omega_2)=0\implies \cos s=\pm\sec\omega_1\sin(\omega_2-\psi) \,,   \label{311b}
\\ 
\sin(\omega_2-\psi)=0\implies \cos s=\pm\cos\omega_1\cos(\omega_0+\psi)\,.     \ \label{311c}
\end{gather}
\end{subequations} 

Equations \eqref{311a} and \eqref{311c}, give rise to an equally spaced spectrum on the variable $s$ (not for the energy), for~which $s=n\pi+f(\omega_0,\omega_1,\psi)$, $n=0,1,2,\dots$. In~any case, the~minimal energy level is given by $f(\omega_0,\omega_1,\psi)$. The~determination of this minimal energy is not a trivial matter for \eqref{311c}, since its solution $s=  \arccos(\cos(\omega_1)\cos(\omega_0+\psi))$ is given by a multi-valued~function. 

Equation \eqref{311b} is even more problematic, as~its right hand side may be bigger than one in modulus. One may think that this formula provides the negative energy values for $|\cos s|>1$. However, we have to keep in mind that there are only possible two negative energy levels, if~any, or~if there is only one, this could be either single or doubly-degenerate, so that \eqref{311b} may not give solutions to the energy spectrum and should be discarded, in~principle.

\subsection{About the Negative and Zero~Energies}

Up to now, we have not been interested in zero and negative values of the energy. Observe that the transcendental equation \eqref{13}, which gives the energy levels, is valid for those extensions, $H_\alpha$,  having positive energies only. These energy levels are, in~all positive energy cases, infinite. 

If we wanted to analyze those Hamiltonians $H_\alpha$ with negative energy levels, we need to perform the replacement  $s\to-i r$ in the wave function \eqref{6} as well as in \eqref{13}. The~latter appears in terms of hyperbolic functions and may have one or two solutions with zero or negative energies. If~there were just one negative energy level, this is doubly degenerate~\cite{NAI}.

When the ground state shows a negative energy, its wave function is similar to \eqref{6}, where the trigonometric functions have been replaced by hyperbolic functions. In~this case, the~ground state wave functions may have zeros (nodes) on the interval $[-a,a]$. Here, the~general formalism says that the procedure to obtain the SUSY partners is not valid~\cite{CHIS}. Nevertheless, this formalism gives a procedure and this procedure may still be applied in this case. The~result is clear---instead of obtaining a new potential with a countable infinite number of equally spaced values of the square root of the eigenvalues of the Hamiltonian ($s$), we obtain new Hamiltonians with either a finite number of eigenvalues or a continuous spectrum only. In~the first case, these energy eigenvalues come from a transcendental equation. In~the second case, partner potentials are often singular, showing an infinite divergence.  We shall discuss this situation in detail in a forthcoming publication. A~similar situation emerges when the ground state has zero~energy. 

From now on, we will concentrate in obtaining the SUSY partners of the self-adjoint extensions that we have analyzed up to~now.

\section{Supersymmetric Partners for the Simplest~Extensions}
\label{sec4}

In this section we shall consider the first and second order supersymmetry transformation applied to some of the self-adjoint extensions $H_\alpha$ so far considered in~here.

\subsection{First Order SUSY~Partners}

The technique to obtain the SUSY partner corresponding to a given self-adjoint operator with discrete spectrum has been discussed in~\cite{CHIS}. To~begin with, let us fix some notation and call $H_\alpha$ to the self-adjoint extension characterized by the values $\alpha:= (m_0,m_1,m_2,m_3,\psi)$ of the~parameters.  

Then, let us follow the procedure of~\cite{CHIS} to obtain the SUSY partners of $H_\alpha$. First of all, we need to determine the ground state $\phi_\alpha^{(0)}(x)$ of $H_\alpha$. This ground state has energy $E_\alpha^{(0)}= (s_\alpha^{(0)}/(2a))^2$, which may be in principle either positive or negative. In the present paper, we shall deal with those extensions having the ground level with positive energy and, for~all  energy levels, $s_\alpha^{(n)}$, $n=0,1,2,\dots$, we have $s_\alpha^{(n)}=(n+1)\,s_\alpha^{(0)}$.

In general, there are two supersymmetric partners of the self-adjoint extension $H_\alpha$, which are  Hamiltonians of the form $-d^2/dx^2 + V^{(j)}_\alpha$ , where $V^{(j)}_\alpha$, $j=1,2$, are a pair of new potentials which is called  partner potentials.  In~order to obtain each of the $V^{(j)}_\alpha$, pick the ground state $\phi_\alpha^{(0)}(x)$ of $H_\alpha$. The~explicit form of this ground state is, after~\eqref{6},
\begin{equation}\label{27}
    \phi_{\alpha}^{(0)}(x)=A(s_0)\cos\left(\frac{s_0}{2a}x\right)+B(s_0)\sin\left(\frac{s_0}{2a}x\right)\,,
\end{equation}
where we have used the simplified notation $s_0:= s_\alpha^{(0)}$, which we shall henceforth keep for simplicity unless otherwise stated. Since \eqref{27} must be in the domain of $H_\alpha$,  the~coefficients $A(s_0)$ and $B(s_0)$ must satisfy the boundary conditions defining this domain. Although~these coefficients depend on the energy ground state $s_0$, we shall also omit this dependence, unless~necessary. Then, we construct the partner potentials $V^{(j)}_\alpha$, $j=1,2$ using an intermediate function called the {\it super-potential}, $W_\alpha(x)$, which is defined as
\begin{equation}\label{33}
W_\alpha(x):=-\dfrac{\partial_x\phi^{(0)}_\alpha(x)}{\phi_\alpha^{(0)}(x)}\,,
\end{equation}
where $\partial_x$ means derivative with respect to $x$. 

Now, we construct the partner potentials $V^{(j)}_\alpha$, $j=1,2$, as~\cite{CHIS}
\begin{eqnarray}\label{34}
   V^{(1)}_\alpha(x)\!\!&\!\!=\!\!&\!\!W_\alpha^2(x)-W_\alpha'(x)=-\left(\frac{s_0}{2a}\right)^2\,, 
   \\[2ex]
 V^{(2)}_\alpha(x)\!\!&\!\!=\!\!&\!\!W_\alpha^2(x)+W'_\alpha(x)=\left(\frac{s_0}{2a}\right)^2\left(1+2\left(\frac{A\sin\left(\frac{s_0}{2a}x\right)-B\cos\left(\frac{s_0}{2a}x\right)}{A\cos\left(\frac{s_0}{2a}x\right)+B\sin\left(\frac{s_0}{2a}x\right)} \right)^2\right)\,.
 \label{3444}
\end{eqnarray}

According to \eqref{34}, it comes that $ V^{(1)}_\alpha(x)$ is constant and equal, in~modulus, to~the original system lowest energy level. We see that this solution is trivial, as~only shifts the energy levels. If~we represent as $\phi^{(1)}(x)$ and $E_n^{(1)}$ the wave function of the ground state and the $n$-th energy level in this situation, we have that
\begin{equation}\label{35}
\phi^{(1)}(x)\equiv \phi^{(0)}(x)\,, \qquad E^{(1)}_n=\left(\frac{s_0}{2a}\right)^2(n^2-1)\equiv E^{(0)}_n-E^{(0)}_{n=1}\,, 
\end{equation}
and $n=1,2,\dots$ is arbitrary. 
In the sequel, we omit the subindex $\alpha$ for simplicity in the notation, unless~otherwise stated for~necessity. 

The Schr\"odinger equation coming from the second potential in \eqref{3444} is
\begin{equation}\label{36}
-\frac{d^2}{dx^2}\phi^{(2)}(x)+\left(\frac{s_0}{2a}\right)^2\left(1+2\left(\frac{A\sin\left(\frac{s_0}{2a}x\right)-B\cos\left(\frac{s_0}{2a}x\right)}{A\cos\left(\frac{s_0}{2a}x\right)+B\sin\left(\frac{s_0}{2a}x\right)} \right)^2\right)\phi^{(2)}(x)=E^{(1)}_n\phi^{(2)}(x)\,,
\end{equation}
where the meaning of $\phi^{(2)}(x)$ is obvious. Next, let us define a new variable $z$ as
\begin{equation}\label{37}
\frac{s_0}{2a}z=iW(x)=\frac{is_0}{2a}\frac{A\sin\left(\frac{s_0}{2a}x\right)-B\cos\left(\frac{s_0}{2a}x\right)}{A\cos\left(\frac{s_0}{2a}x\right)+B\sin\left(\frac{s_0}{2a}x\right)} \,.
\end{equation}

Under this change of variables, the~Schr\"odinger Equation \eqref{36} takes the form:
\begin{equation}\label{38}
    (1-z^2)\, \partial_z^2 {\phi}^{(2)}(z)-2z\, \partial_z {\phi}^{(2)}(z)+\left(\ell(\ell+1)-\frac{n^2}{1-z^2}\right){\phi}^{(2)}(z)=0\,, \quad \text{with}\quad \ell=1,
\end{equation}
where $\partial_z$ represents the derivation with respect to $z$. This is a particular case of associated Legendre equation when $\ell=1$, and~their solutions are well known. One of them is given by the associated Legendre functions of second kind:
\begin{equation}\label{39}
Q_\ell^n(z):= (-1)^n\, (1-z^2)^{n/2}\, \frac{d^n}{dz^n}\,Q_\ell(z)\,,
\end{equation}
where $Q_\ell(z)$ are the Legendre functions of the second kind~\cite{AS}. These solutions for \eqref{38} provide the solutions for \eqref{36}:
\begin{equation}\label{40}
    \phi^{(2)}_n(x)=Q_1^n\left(i\frac{A\sin\left(\frac{s_0}{2a}x\right)-B\cos\left(\frac{s_0}{2a}x\right)}{A\cos\left(\frac{s_0}{2a}x\right)+B\sin\left(\frac{s_0}{2a}x\right)} \right)\,, 
\end{equation}
where, of~course, we have taken the value $\ell=1$. 

In addition, there is another set of solutions given by the first kind associated Legendre functions $P^n_1(z)$. These functions have not been considered as solutions to our problem, since they show singularities within the open interval $(-a,a)$ and are not square integrable, as~shown in Appendix~\ref{apendice}. In~Figure~\ref{fig:fig3}, we represent some of the wave equations just obtained in \eqref{40} for the lowest energy levels. 
Let us consider now the {\it {second partner Hamiltonian}}, $H_\alpha^{(2)} : = H_\alpha +V_\alpha^{(2)}$, or~$H^{(2)}$ in brief. The~solution $Q_1^1(z)$ shows a logarithmic singularity at each of its extremes and, therefore, it is not square integrable. Nevertheless, for~$n \ge 2$, these solutions are square integrable, as is proved in the~{Appendix} \ref{apendice}.

\begin{figure}[htb]
        \includegraphics[width=0.85\textwidth]{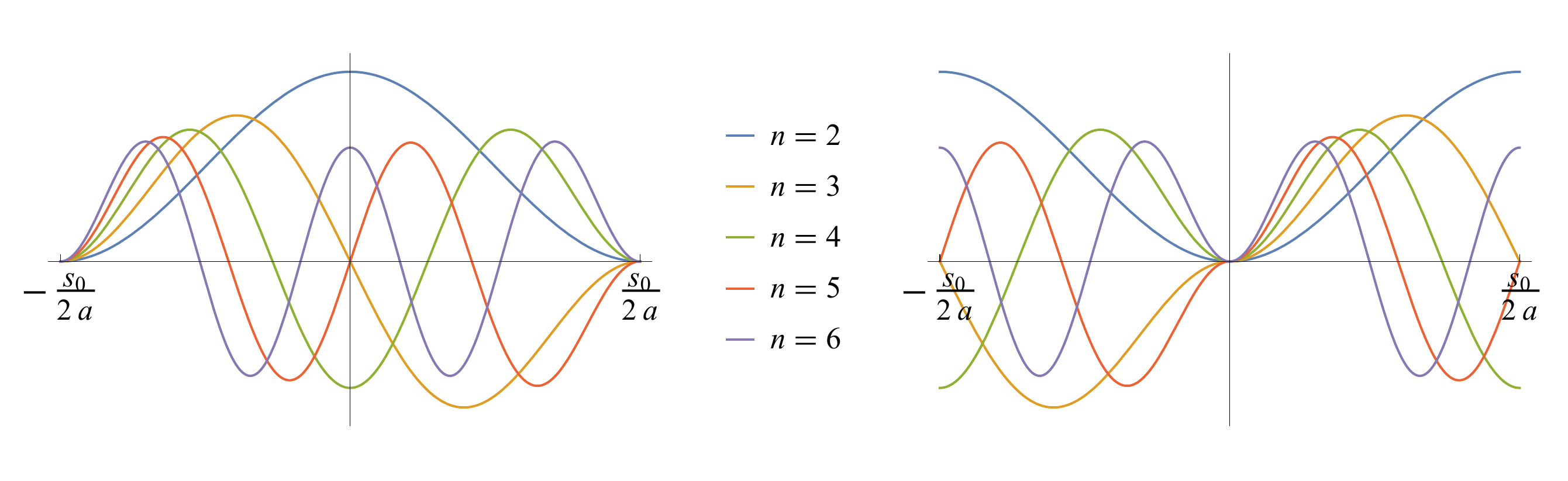}
\caption{First order supersymmetric (SUSY) states $\phi^{(2)}_n(x)$ from \eqref{40} when the ground state of the original system is either purely even, that is $B=0$ (plot on the left), or~purely odd, that is $A=0$ (plot on the right).
Note that the quantum number  $n$ of the Legendre function in  \eqref{40} is the number of the nodes of the function.}
\label{fig:fig3}
\end{figure}

\subsection{Second Order SUSY~Partners}

Once we have obtained the first order SUSY partners for the self-adjoint extensions $H_\alpha$ with ground state of positive energy, let us inspect how we may obtain an infinite chain of higher order partners for $H_\alpha$. In~the all above discussed cases, the~bound state has wave function given by $Q_1^2(z)$ (which is obviously not the same for all cases, since the definition of $z$ changes). Then,  we obtain the super-potential $W^{(2)}(x)$ by replacing $\phi^{(0)}_{n=1}(x)$ by $Q_1^2(z)$ (and then write $z$ in terms of $x$) in \eqref{33}. This procedure gives rise to two second order potential partner candidates, which are:
\begin{eqnarray}
V^{(2,1)}(x)\!\!&\!\!=\!\!&\!\! W_{(2)}^2(x)-W_{(2)}'(x)=\left(\frac{s_0}{2a}\right)^2\left(1+2\left(\frac{A\sin\left(\frac{s_0}{2a}x\right)-B\cos\left(\frac{s_0}{2a}x\right)}{A\cos\left(\frac{s_0}{2a}x\right)+B\sin\left(\frac{s_0}{2a}x\right)} \right)^2\right)-3\left(\frac{s_0}{2a}\right)^2\,, \label{41} 
\\[1ex]  
V^{(2,2)}(x)\!\!&\!\!=\!\!&\!\!W_{(2)}^2(x)+W_{(2)}'(x)=\left(\frac{s_0}{2a}\right)^2\left(1+3\left(\frac{A\sin\left(\frac{s_0}{2a}x\right)-B\cos\left(\frac{s_0}{2a}x\right)}{A\cos\left(\frac{s_0}{2a}x\right)+B\sin\left(\frac{s_0}{2a}x\right)} \right)^2\right)\,. \label{42}
\end{eqnarray}

Although the notation used in \eqref{41} and \eqref{42} should be clear, we need a generalization of it, as~we are going to consider further order partners next. Thus, we shall use $V^{(i,j)}(x)$ and $W_{(i)}(x)$, where the index $i$ gives the order of the partner, which in the above case is $i=2$. This index may take all possible values $i=1,2,3,\dots$ The index $j$ always takes two possible values, $j=1,2$. From~this point of view, $V_\alpha^{(i)}(x)$ in \eqref{34} could  be written as $V^{(1,i)}(x)$. Analogously, we may use for the $i$-th partner Hamiltonian the notation $H^{(i,j)}$. To~simplify the notation, we have always omitted the subindex $\alpha$, which labels the precise self-adjoint extension we are~considering. 

Observe that according to \eqref{34} and \eqref{41}, $V^{(2,1)}(x)= V^{(1,2)}(x)-3(s_0/(2a))^2=V^{(2)}_\alpha -3(s_0/(2a))^2$, so that $H^{(2,1)}$ and $H^{(1,2)}$ have the same eigenvalues shifted by $3(s_0/(2a))^2$. Thus, we ignore \eqref{41} and solely consider \eqref{42}. For~\eqref{42}, we may do a similar analysis than in the previous case, that is,~ first
order SUSY partner, so that the bound state wave functions are given by
\begin{equation}\label{43}
 \phi^{(2,2)}_n(x)=Q^n_{2}\left(i\frac{A\sin\left(\frac{s_0}{2a}x\right)-B\cos\left(\frac{s_0}{2a}x\right)}{A\cos\left(\frac{s_0}{2a}x\right)+B\sin\left(\frac{s_0}{2a}x\right)} \right)\,.
\end{equation}

\begin{figure}[htb]
   \includegraphics[width=0.85\textwidth]{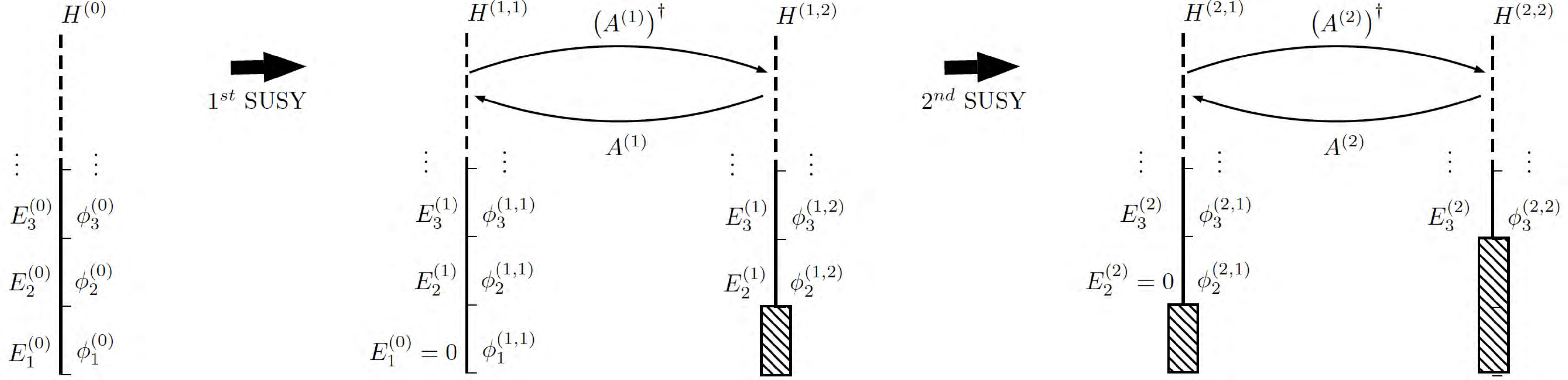}
\caption{Different energy levels of first and second supersymmetry~Hamiltonians.}
\label{fig:fig4}
\end{figure}

In this second order SUSY, both functions $Q_2^1(z)$ and $Q_2^2(z)$ have logarithmic singularities at the points $x=\pm a$, so that they are not square integrable on $[-a,a]$ and, consequently, should be discarded as proper eigenfunctions of $H^{(2,2)}$. Thus, the~ground state for $H^{(2,2)}$ has a wave function given by $Q_2^3(z)$. This is a general behaviour that could be checked at each step going from a SUSY partner to the next one, as is shown in  Figure~\ref{fig:fig4}.

\section{Supersymmetric Self-Adjoint Extensions of the Infinite Well at \boldmath$\ell$-Order}
\label{sec5}

Let us begin this Section with a summary of the notation employed so far and its~meaning:
\begin{equation*}
    \begin{cases}
    H_\alpha & \text{Original Hamiltonian, which is a self-adjoint extension of $H_0=-d^2/dx^2$. }\\
    \phi^{(0)}_n  & \text{Wave function of $H_\alpha$ associated to the $n-$level.}\\
    E^{(0)}_n  & \text{Energy spectrum of $H_\alpha$.}\\
    \phi^{(i,1)}_n  &\text{Wave function of first SUSY partner at $i$ order associated to the $n-$ level.}\\
    \phi^{(i,2)}_n & \text{Wave function of second SUSY partner at $i$ order associated to the $n-$ level.}\\
    W_{(i)}  &\text{Super potential at $i$ order, calculated from the second partner wave function.}\\
     &\text{of previous SUSY order, that is,~$ \phi^{(i-1,2)}_{i}$.}\\
    V^{(i,1)},V^{(i,2)}   & \text{Partner potentials of  $i-$order SUSY constructed from $W_{(i)}$.}\\
     A^{(i)}, (A^{(i)})^\dagger   & \text{Annihilation/Creation operator of SUSY at $i-$order.}\\
    \end{cases}
\end{equation*}

Creation $(A^{(i)})^\dagger$ and annihilation $A^{(i)}$ operators will be defined~later. 

So far, we have obtained potentials and wave functions for the first and second SUSY partners for self-adjoint extensions of $H_0=-d^2/dx^2$ with ground level of positive energy. With~the help of the induction method, we may find potentials as well as wave functions and energy levels for arbitrary order $\ell$ SUSY partners for the same class of self-adjoint extensions. We have seen already that from the SUSY partners $V^{(i,1)},V^{(i,2)}$, only the last one  is really interesting and we will focus on it in the~sequel.

In order to apply the inductive method, let us assume that the ground state for the $\ell$-th SUSY partner, $H^{(\ell,2)}$, of~$H_\alpha$ is given by
\begin{equation}\label{44}
  \phi^{(\ell,2)}_{\ell+1}(x)= Q^{\ell+1}_{\ell}\left(i\frac{A\sin\left(\frac{s_0}{2a}x\right)-B\cos\left(\frac{s_0}{2a}x\right)}{A\cos\left(\frac{s_0}{2a}x\right)+B\sin\left(\frac{s_0}{2a}x\right)} \right),
\end{equation}
as in the previous cases \eqref{40} and \eqref{43}.
Then, the~super-potential takes the following form:
\begin{equation}\label{45}
    W_{(\ell+1)}=-\frac{\partial_x^2\phi^{(\ell,2)}(x)}{\phi^{(\ell,2)}(x)}=\frac{s_0 (\ell+1)}{2a}\left(\frac{A\sin\left(\frac{s_0}{2a}x\right)-B\cos\left(\frac{s_0}{2a}x\right)}{A\cos\left(\frac{s_0}{2a}x\right)+B\sin\left(\frac{s_0}{2a}x\right)} \right)=-i\frac{s_0 (\ell+1)}{2a}z\,,
\end{equation}
where $\partial_x^2$ denotes the second derivative with respect to the variable $x$. Once we have the { super potential}, we readily obtain the partner potentials at $\ell+1$ order, which are
\begin{eqnarray}
V^{(\ell+1,1)}(x)\!&\!=\!&\!( W_{(\ell+1)})^2-\partial_x  W_{(\ell+1)}=\frac{s_0^2(\ell+1)}{(2a)^2}\left(-1+\ell\left(\frac{A\sin\left(\frac{s_0}{2a}x\right)-B\cos\left(\frac{s_0}{2a}x\right)}{A\cos\left(\frac{s_0}{2a}x\right)+B\sin\left(\frac{s_0}{2a}x\right)}\right)^2 \right)\,, \label{46} 
\\ [1ex]   
V^{(\ell+1,2)}(x)\!&\!=\!&\!( W_{(\ell+1)})^2+\partial_x  W_{(\ell+1)}=\frac{s_0^2(\ell+1)}{(2a)^2}\left(1+(\ell+2)\left(\frac{A\sin\left(\frac{s_0}{2a}x\right)-B\cos\left(\frac{s_0}{2a}x\right)}{A\cos\left(\frac{s_0}{2a}x\right)+B\sin\left(\frac{s_0}{2a}x\right)}\right)^2 \right)\,.\label{47}
\end{eqnarray}

Note that although the label $\alpha$ is not written explicitly on the above equations and many others, potentials and wave functions must depend on $\alpha$. This dependence is hidden in $s_0$, where we have not made it explicitly for simplicity in the~notation.  

The Schr\"odinger equation for the first $(\ell+1)$-th order partner potential, $V^{(\ell+1,1)}$,  is
\begin{equation}\label{48}
    -\partial_x^2\phi^{(\ell+1,1)}(x)+V^{(\ell+1,1)}(x)\phi^{(\ell+1,1)}(x)=E^{(\ell+1)}\phi^{(\ell+1,1)}(x)\,.
\end{equation}

If we change it to the $z$ variable, \eqref{48} takes the form:
\begin{eqnarray}\label{49}
(1-z^2)^2\partial_z^2\phi^{(\ell+1,1)}(z)-2z(1-z^2)\partial_z\phi^{(\ell+1,1)}(z)-(\ell+1)(1+\ell z^2)\phi^{(\ell+1,1)}(x) =
\left(\frac{2a}{s_0}\right)^2E^{(\ell+1)}\phi^{(\ell+1,1)}(x)\,. 
\end{eqnarray}

Equation \eqref{49} is a new Legendre-type equation for which solutions are known. The~respective eigenfunctions and eigenvalues are
\begin{equation}\label{50}
\phi^{(\ell+1,1)}_n(x)=Q^{n}_{\ell}\left(i\frac{A\sin\left(\frac{s_0}{2a}x\right)-B\cos\left(\frac{s_0}{2a}x\right)}{A\cos\left(\frac{s_0}{2a}x\right)+B\sin\left(\frac{s_0}{2a}x\right)} \right)\,, \qquad E^{(\ell+1)}_n=\left(\frac{s_0}{2a}\right)^2\left(n^2-(\ell+1)^2\right)\,.
\end{equation}

Observe that the first partner wave functions of order $\ell+1$ are the same as the second partner wave functions of order $\ell$, that is,~$\phi^{(\ell+1,1)}_n(x)\equiv \phi^{(\ell,2)}_n(x)$. The~Schr\"odinger equation with potential $V^{(\ell+1,2)}(x)$ is
\begin{equation}\label{51}
    -\partial_x^2\phi^{(\ell+1,2)}(x)+V^{(\ell+1,2)}(x)\phi^{(\ell+1,2)}(x)=E^{(\ell+1)}_n\phi^{(\ell+1,2)}(x)\,,
\end{equation}
which in terms of the $z$ variable becomes:
\begin{eqnarray}\label{5222}
(1-z^2)^2\partial_z^2\phi^{(\ell+1,2)}(z)-2z(1-z^2)\partial_z\phi^{(\ell+1,2)}(z)+ (\ell+1)(1-(\ell+2)z^2)\phi^{(\ell+1,2)}(x) =\left(\frac{2a}{s_0}\right)^2E^{(\ell+1)}_n\phi^{(\ell+1,2)}(x)\,.
\end{eqnarray}

Equation \eqref{5222} is again of Legendre type and its solutions in terms of eigenfunctions have the form:
\begin{equation}\label{53}
 \phi^{(\ell+1,2)}_n(x)=Q^{n}_{\ell+1}\left(i\frac{A\sin\left(\frac{s_0}{2a}x\right)-B\cos\left(\frac{s_0}{2a}x\right)}{A\cos\left(\frac{s_0}{2a}x\right)+B\sin\left(\frac{s_0}{2a}x\right)} \right)\,.
\end{equation}

The energy spectrum is given by
\begin{equation}\label{54}
E^{(\ell+1)}_n=\left(\frac{s_0}{2a}\right)^2(n^2-(\ell+1)^2).
\end{equation}

Finally, one defines the annihilation, $ A^{(\ell+1)}$, and~creation, $(A^{(\ell+1)})^\dagger$ operators, which transform the eigenvectors of $H^{(\ell+1,1)}$ into the eigenvectors of $H^{(\ell+1,2)}$ and reciprocally, respectively, as:
\begin{eqnarray}
 A^{(\ell+1)} \!&\!=\!&\! \partial_x+W^{(\ell+1)}(x)=\frac{i s_0}{2a}(1-z^2)\partial_z-i(\ell+1)\frac{s_0}{2a}z\,, \label{55} \\ [1ex] 
 (A^{(\ell+1)})^\dagger \!&\!=\!&\! -\partial_x+W^{(\ell+1)}(x)=-\frac{i s_0}{2a}(1-z^2)\partial_z-i(\ell+1)\frac{s_0}{2a}z\,.
\end{eqnarray}

These creation and annihilation operators have been already constructed for the general formalism of SUSY potential partners in~\cite{CHIS}. 

\begin{figure}[htb]
   \includegraphics[width=0.85\textwidth]{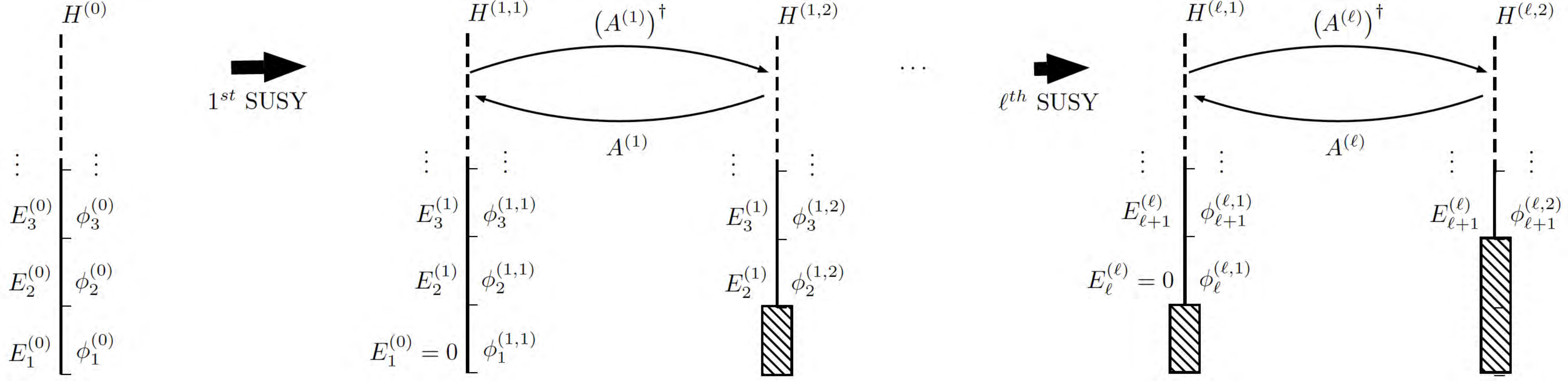}
\caption{Energy scheme of different SUSY transformations up to order $\ell$.}
\label{fig:fig5}
\end{figure}

The relation between the Hamiltonian partners $H^{(\ell,1)}$ and $H^{(\ell,2)}$ for $\ell$ arbitrary are shown in  Figure~\ref{fig:fig5}. For~$\ell=0$, there is a unique Hamiltonian, which is $H_\alpha$. Now, the~creation and annihilation operators in the $z$ variable give the recurrence identities for the associated Legendre~functions.

\section{Conclusions and~Outlook}
\label{sec6}

We have discussed  the results of~\cite{BFV} for the self-adjoint extensions of the differential operator $H_0=-d^2/dx^2$ and gone beyond these results in the sense of addressing some cases not treated in~\cite{BFV}. Also, we have proposed a more detailed classification of the spectrum of these extensions in terms of the parameters that characterize each one of these extensions. We have seen that it is possible to classify these extensions in terms of other sets of variables with the sense of angles, which permits us to go beyond~\cite{BFV}.  These self-adjoint extensions may have at most two negative eigenvalues, a~ground state of zero energy and ground states with strictly positive~energy. 

In addition, in~this paper, we have obtained analytically the form of the SUSY partners for the self-adjoint extensions of $H_0$ (that we denote as $H_\alpha$, where $\alpha$  includes the four real parameters that gives each of these extensions) with a ground state with positive energy.  We have obtained all Hamiltonian partners of each of the $H_\alpha$ with positive spectrum to all orders, their energy levels and their eigenfunctions. At~each step, we find two distinct Hamiltonian partners of $\ell$-th order. Creation and annihilation operators related the eigenfunctions for these two partners were also~evaluated. 

Although we have obtained the eigenfunctions for the whole sequence of SUSY partners of each of the $H_\alpha$, these eigenfunctions depend explicitly on the square root of the ground state energy of $H_\alpha$, which in most cases can be obtained by solving a transcendental equation. However, this transcendental equation looks rather intractable in a few cases. This situation poses some difficulties in obtaining the eigenvalues for some of the $H_\alpha$, although~the explicit form of their eigenfunctions and of the eigenfunctions of their SUSY partners can always be given, as~functions of the square root of the ground state energy of~$H_\alpha$.  

We have not obtained the SUSY partners for those extensions, $H_\alpha$, with~a ground state with zero or negative energy. Here, we may also obtain a sequence of SUSY partners form each of the $H_\alpha$ in this class. Unlike the partners for extensions $H_\alpha$ with ground states with strictly positive energies, these partners may have a finite number of eigenvalues or even none, and the potential partners may show singularities. A~classification of the partners for these exceptional extensions is left for a forthcoming~paper.

\acknowledgements

This research was funded by Junta de Castilla y Le\'on and FEDER projects VA137G18 and~BU229P18.

\appendix

\section{}
\label{apendice}

In this Appendix, we justify the correct choice of the wave functions for the bound states of the supersymmetric partners of each of the extensions $H_\alpha$ with strictly positive ground state energy. In~Appendix~\ref{a1}, we derive a general solution for these wave functions  as a linear combination of the associated Legendre functions $P_\ell^n$ and $Q_\ell^n$ with argument $-\tan((s_0x)/(2a))$. In~Appendix~\ref{a2}, we show that the component with $P_\ell^n$ should be discarded, since it does not meet the requirement of square integrability. On~the other hand, the~component with $Q_\ell^n$ should give the wave function as is square integrable, as~proven in Appendix~\ref{a3}.

\subsection{General solution for the wave functions }
\label{a1}

Comments in these Appendices are valid for those self-adjoint extensions $H_\alpha$ with ground states with positive energy. For~each of these extensions, the~ground state energy is $E_0^{(0)}= (s_0/(2a))^2$, where $s_0$ depends on the chosen self-adjoint extensions and, therefore, on~the values of the parameters. 
As we have seen, in~terms of the auxiliary variable $s$, the~spectrum is equally spaced in this case, so that all other energy values are  $E_m^{(0)}=(s_0/(2a))^2m^2$. For~the ground state, the~wave function is
\begin{equation}\label{52}
    \phi^{(0)}_{m=1}(x)=A\cos\left(\frac{s_0}{2a}x\right)+B\sin\left(\frac{s_0}{2a}x\right)\,.
\end{equation}

The coefficients $A$ and $B$, as~complex numbers,  should have the same phase in order to have a real partner potential. To~see it, let us write $A=C e^{i\varphi_1}$ and $B=D e^{i\varphi_2}$, with~$C:=|A|$ and $D:=|B|$. Then, \eqref{52} is
\begin{equation}\label{533}
   \phi^{(0)}_{m=1}(x)=  C e^{i\varphi_1}\cos\left(\frac{s_0}{2a}x\right)+D e^{i\varphi_2}\sin\left(\frac{s_0}{2a}x\right)\,.
\end{equation}

Using Definitions \eqref{41} and \eqref{42} for the potential partners of $\ell$-th order, we have for the first $\ell$-th partner:
\begin{equation}\label{544}
 V^{(\ell+1,1)}=\frac{s_0^2(\ell+1)}{(2a)^2}\left(-1+\ell\left(\frac{C e^{i\varphi_1}\sin\left(\frac{s_0}{2a}x\right)-D e^{i\varphi_2}\cos\left(\frac{s_0}{2a}x\right)}{C e^{i\varphi_1}\cos\left(\frac{s_0}{2a}x\right)+De^{i\varphi_2}\sin\left(\frac{s_0}{2a}x\right)}\right)^2 \right)\,,
\end{equation}
for which the imaginary part is given by
\begin{eqnarray}\label{555}
 \operatorname{Im}\left(V^{(\ell+1,1)}\right) 
 \! &\!=\! &\!
 \operatorname{Im}\left(\frac{s_0^2(\ell+1)}{(2a)^2}\left(-1+\ell\left(\frac{A\sin\left(\frac{s_0}{2a}x\right)-B\cos\left(\frac{s_0}{2a}x\right)}{A\cos\left(\frac{s_0}{2a}x\right)+B\sin\left(\frac{s_0}{2a}x\right)}\right)^2 \right)\right)  \nonumber\\ [2ex] 
\! &\! =\! &\!
 \frac{C D \ell (\ell+1) s_0^2  \left((C-D)
   (C+D) \sin \left(\frac{s_0 x}{a}\right)-2 C D \cos (\varphi_1-\varphi_2) \cos \left(\frac{s_0 x}{a}\right)\right)}{a^2 \left(2 C
   D \cos (\varphi_1-\varphi_2) \sin \left(\frac{s_0
   x}{a}\right)+(C-D) (C+D) \cos \left(\frac{s_0 x}{a}\right)+C^2+D^2\right)^2}\sin (\varphi_1-\varphi_2)\,,
\end{eqnarray}
so that  potential \eqref{544} is real if $\sin (\varphi_1-\varphi_2)=0$, or~equivalently, if~$\varphi_1=n \pi+\varphi_2$. Thus, if~$A$ and $B$ have the same phase as complex numbers, we have guaranteed that the potential partner $V^{(\ell+1,1)}$ is real. The~same is valid for $V^{(\ell+1,2)}$. Thus, \eqref{533} becomes:
\begin{equation}\label{56}
     \phi^{(0)}_{m=1}(x)=Ce^{i\varphi}\cos\left(\frac{s_0}{2a}x\right)+De^{i\varphi}\sin\left(\frac{s_0}{2a}x\right)\,.
\end{equation}

This ground state is not yet normalized. Its normalization gives
{\small \begin{eqnarray}\label{57}
    \int_{-a}^{a}dx \,\phi_{m=1}^{(0)}(x) \left( \phi_{m=1}^{(0)}(x)\right)^*=1\quad \implies\quad  C^2 +D^2=1 \quad \implies\quad
C=\cos\delta,\ D=\sin\delta \,.
\end{eqnarray}}

Finally, the~ground state wave function has the form:
\begin{equation}\label{58}
    \phi^{(0)}_{m=1}(x)=e^{i\varphi}\cos\left(\frac{s_0}{2a}x+\delta\right)\,.
\end{equation}

Let us recall that our goal is to show that the solution of the Schr\"odinger equation with component $Q^n_\ell$ is square integrable and  the solution with $P^n_\ell$ is not. To~begin with, let us define a new independent variable using the shift $x= y-{2a\delta}/{s_0}$. The~ground state has now the form,
\begin{equation}\label{59}
    \phi^{(0)}_{m=1}=e^{i\varphi}\cos\left(\frac{s_0}{2a}y\right)\,.
\end{equation}

With this notation, the~wave function of the second partner of $\ell$-th order is
\begin{equation}\label{60}
    \phi^{(\ell,2)}_m =C_1 P^n_\ell\left(-i \tan\left(\frac{s_0 y}{2a}\right)\right)+C_2 Q^n_\ell\left(-i \tan\left(\frac{s_0 y}{2a}\right)\right)\,.
\end{equation}
Next, we shall analyze the square integrability of each of the components in \eqref{60}.

\subsection{Trigonometric Expansion of $P^n_\ell\left(-i \tan\left(\frac{s_0 y}{2a}\right)\right)$}
\label{a2}

Let us use the change of variable $z=-i\tan\left(\frac{s_0 y}{2a}\right)$ and consider the hypergeometric form of the associated Legendre functions  with argument $z$  \cite{AS}:
\begin{eqnarray}
 P_\ell^n(z) \!&\!=\!&\! \frac{1}{\ell!}\left(-\frac{1}{2}\right)^\ell \left(\frac{1+z}{1-z}\right)^{n/2} (1-z)^{\ell}
  \frac{\Gamma (2 \ell+1) \, _2F_1\left(-\ell,n-\ell;-2 \ell;-\frac{2}{z-1}\right)}{\Gamma (\ell-n+1)} 
  \nonumber\\
   \!&\!=\!&\!  \frac{1}{\ell!}\left(-\frac{1}{2}\right)^\ell \left(\frac{1+z}{1-z}\right)^{n/2} (1-z)^{\ell}
  \sum _{j=0}^\ell \frac{ \left((2 \ell-j)!
   (-\ell)_j\right)}{j! \Gamma (-j+\ell-n+1)}\left(\frac{2}{1-z}\right)^j 
   \nonumber \\
   \!&\!=\!&\! \frac{\left(-\frac{1}{2}\right)^\ell \Gamma (2 \ell+1)  }{\ell! \Gamma (\ell-n+1)}\frac{e^{-\frac{is_0 (n-\ell) }{2 a}y}}{\cos^\ell\left(\frac{s_0  y}{2 a}\right)}\sum_{j=0}^\ell \frac{(-1)^j \Gamma (\ell+1) (2 \ell-j)!}{j! \Gamma (-j+\ell+1) \Gamma (-j+\ell-n+1)}\left(2e^{-\frac{is_0}{2a}y}\cos\left(\frac{s_0  y}{2 a}\right)\right)^j \nonumber \\   
   \!&\!=\!&\!\sum_{j=0}^\ell\frac{(-1)^{j+\ell} 2^{j-\ell} \Gamma (\ell+1) (2 \ell-j)!}{j! \ell! \Gamma (-j+\ell+1) \Gamma (-j+\ell-n
   +1)}e^{-\frac{i s_0  y (j-\ell+n)}{2 a}} \cos ^{j-\ell}\left(\frac{s_0  y}{2 a}\right)\,. 
   \label{a10a}
\end{eqnarray}
Due to the presence of negative powers of the cosine in \eqref{a10a}, the~resulting wave function is not square integrable and, therefore, not acceptable as a wave function of a bound~state.

\subsection{Trigonometric Expansion of $Q^{n}_\ell(-i\tan\left(\frac{s_0 y}{2a}\right))$}
\label{a3}

Similarly, we can express $Q^{n}_\ell(z)$ in terms of a hypergeometric function~\cite{AS} as:
\begin{equation*}
 Q^n_\ell(z)=  \frac{1 }{\sqrt{\pi }}2^{-\ell-1} (-1)^{\ell+n+1} (z-1)^{-\ell-1} \Gamma \left(-\ell-\frac{1}{2}\right)
   \left(\frac{z+1}{z-1}\right)^{n/2} (\ell+n)!
   \ _2F_1\left(\ell+1,\ell+n+1;2
   (\ell+1);-\frac{2}{z-1}\right)\,.
\end{equation*}

Then, let us perform again the change of variables given by $z=-i\tan\left(\frac{s_0 y}{2a}\right)$, so as to~obtain:
\begin{eqnarray*}
 Q^n_\ell\left(-i\tan\left(\frac{s_0 y}{2a}\right)\right) \!\!&\!\! =  \!\!&\!\!  \frac{1}{\sqrt{\pi }}2^{-\ell-1} (-1)^n \Gamma \left(-\ell-\frac{1}{2}\right) \Gamma (\ell+n+1) e^{-i \frac{s_0}{2a} y
   (\ell+n+1)} \cos ^{\ell+1}\left(\frac{s_0 y}{2a}\right)
\\
    \!\!&\!\!   \!\!&\!\!  \times \, _2F_1\left(\ell+1,\ell+n+1;2(\ell+1);2e^{ -i \frac{s_0y}{2a}}\cos\left(\frac{s_0 y}{2a}\right)\right)\,.
\end{eqnarray*}

If again, we perform a series expansion around $z=0$ we obtain the following power series in terms of positive powers of cosines:
\begin{equation*}
 Q^n_\ell\left(-i \tan\left(\frac{s_0 y}{2a}\right)\right)=  
 \frac{1}{2} (-1)^{-\ell+n+1} e^{-i \frac{s_0}{2a}  n y} \Gamma (n-\ell) \Gamma (\ell+n+1)
    \sum_{j=\ell+1}^{n}  \frac{(-1)^j  \Gamma (j)\ 2^{j}e^{i j \frac{s_0}{2a}  y} }{\Gamma (j-\ell)
   \Gamma (j+\ell+1) \Gamma (-j+n+1)} \cos ^j\left(\frac{s_0}{2a}  y\right)\,.
\end{equation*}
This solution is acceptable as is square~integrable.

\end{document}